# Lumped Element Kinetic Inductance Detectors for space applications

Alessandro Monfardini[*a], Jochem Baselmans[b], Alain Benoit[a], Aurelien Bideaud[a], Olivier Bourrion[c], Andrea Catalano[c,a], Martino Calvo[a], Antonio D'Addabbo[d,a], Simon Doyle[e], Johannes Goupy[a], Helene Le Sueur[f], Juan Macias-Perez[c], for the NIKA2, SPACEKIDS and B-SIDE collaborations

[a]Institut Néel, CNRS and Université de Grenoble, 25 rue des Martyrs, Grenoble, France;
[b]SRON, Sorbonnelaan 2, Utrecht, Netherlands;
[c]LPSC, Laboratoire de Physique Subatomique et Cosmologie, 53 rue des Martyrs, Grenoble, France;
[d]INFN, Laboratori Nazionali del Gran Sasso, Assergi (AQ), Italy;
[e]University of Cardiff, School of Physics and Astronomy, The Parade, Cardiff, UK;
[f]CSNSM, Centre de Sciences Nucléaires et de Sciences de la Matière, Orsay Campus, France

## ABSTRACT

Kinetic Inductance Detectors (KID) are now routinely used in ground-based telescopes. Large arrays, deployed in formats up to kilopixels, exhibit state-of-the-art performance at millimeter (e.g. 120-300 GHz, NIKA and NIKA2 on the IRAM 30-meters) and sub-millimeter (e.g. 350-850 GHz AMKID on APEX) wavelengths. In view of future utilizations above the atmosphere, we have studied in detail the interaction of ionizing particles with LEKID (Lumped Element KID) arrays. We have constructed a dedicated cryogenic setup that allows to reproduce the typical observing conditions of a space-borne observatory. We will report the details and conclusions from a number of measurements. We give a brief description of our short term project, consisting in flying LEKID on a stratospheric balloon named B-SIDE.

## 1. INTRODUCTION

LEKID (Lumped Element Kinetic Inductance Detectors) are a particular implementation of the more general KID (Kinetic Inductance Detectors), based on planar superconducting resonators. In a LEKID, the inductive and capacitive parts of the resonator are separated by design. This trick allows shaping of the inductor, in which the current distribution is at first order homogeneous, to directly absorb the incoming radiation [1]. For all the applications and measurements described in this paper, the detectors consist in a single, patterned, thin (< 35 nm) film deposited on a common Silicon substrate with rather standard thickness (e.g. 150-500 μm). The metal filling factor is in general quite small, of the order of a few percents and up to 10 %.

The NIKA instrument, operated at the 30-meters IRAM telescope at Pico Veleta until 2015, represented a pathfinder for LEKID technology applied to millimeter and sub-millimeter astronomy. The performances, in terms of sensitivity and photometry, have been well in line with the state-of-the art of competing technologies [2,3]. The NIKA2 camera, operational since October 2015 [4,5] and employing 3,300 pixels, demonstrated a relatively safe extension to ten times bigger arrays formats.

In order to fill the gap between the technology demonstrated on ground-based instruments and future space missions, e.g. CORE+ or LiteBird concerning CMB (Cosmic Microwave Background), we are working on a balloon program named B-SIDE that will employ LEKID to map the interstellar dust polarization at frequencies comprised between 400 and 700 GHz. This experience, besides intrinsic scientific implications, will allow us to gain further confidence and to validate the technology under realistic conditions.

In this framework we have studied in detail the behavior of LEKID arrays, with pixels count of the order of 100, under irradiation. We demonstrate that such arrays, and larger ones, are already compatible with space environment as they are. To further reduce the data loss fraction due to the effect of cosmic rays impacts, we have elaborated and demonstrated a recipe employing a Titanium absorber able to confine the particle impact effect on a smaller region of the array.

---

[*] monfardini@neel.cnrs.fr; phone +33 4 76 88 10 52; fax +33 4 56 38 70 87; http://neel.cnrs.fr

## 2. PERFORMANCE UNDER TYPICAL SPACE CONDITIONS

In a previous paper, we have demonstrated that the arrays designed and fabricated for the NIKA project, i.e. assuming optical load per pixel of the order of 5÷50 pW, behave well under the sub-pW load, which is typical of CMB observations from space [6]. Optical sensitivities, measured in real, fully-multiplexed conditions and on 132 pixels arrays, approach photon-noise limitation in the whole band of CMB interest 80÷180 GHz. The sub-band 120÷180 GHz can be covered using standard Aluminum LEKID, while at lower frequency we have proposed, realized and tested bi-layers Ti-Al LEKID [7].

### 2.1 Sensitivity

In large ground-based instruments operating at millimeter wavelengths, the typical load per pixel is of the order of 5÷50 pW, highly dependent upon atmospheric conditions and details of the instrument itself. This lead to target NEP (Noise Equivalent Power) sensitivities of the order of $5 \div 20 \cdot 10^{-17}$ W/Hz$^{0.5}$. Above the atmosphere, a millimeter wave instrument is, on the other hand, loaded mostly by the 2.7 K CMB background. The power hitting a pixel, for typical configurations (e.g. Planck, CORE+, LiteBird), is comprised in the range 0.1÷1 pW depending, among other things, on the bandwidth chosen. The NEP sensitivities that we want to obtain are thus at least one order of magnitude lower compared to the ground-based case. Even before any further design optimization, we have demonstrated in [6] sensitivities in the $1 \div 4 \cdot 10^{-17}$ W/Hz$^{0.5}$ range under 0.3÷0.5 pW load per pixel. The NIKA pixel design was adopted, and two arrays were realized one in Aluminum (18 nm) to cover the band 120÷180 GHz and a second based on a Ti-Al bi-layer ($T_c$ = 0.9 K) to handle the portion 80÷120 GHz.

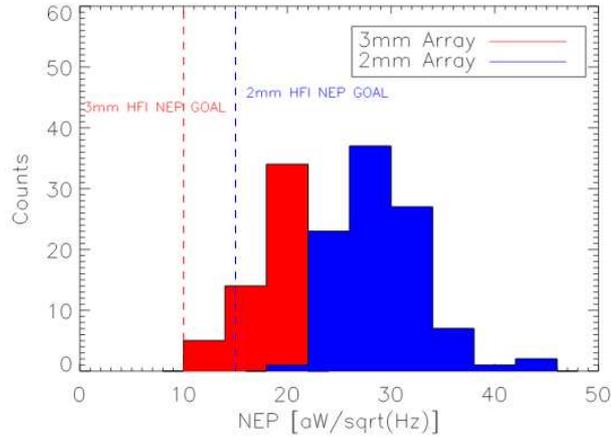

Fig. 1. Distribution of the pixel optical sensitivities (red for the Ti-Al array operating at 80÷120 GHz, blue for the Al array operating at 120÷180 GHz) compared to the reference goals of the Planck HFI detectors. The background per pixel (0.3÷0.5 pW) is representative of the real Planck situation. Taken from [6].

The gain in sensitivity of the NIKA pixels placed under lower background is explained by the natural increase of the quasi-particles recombination time and the better internal quality factor of the resonators under reduced load conditions. The remaining factor to be shaved in order to meet the ultimate goals of the next generation CMB-oriented satellites will be achieved by better matching the electrical parameters of the resonators (e.g. coupling quality factor, kinetic inductance) to the incoming power.

**2.2 Cosmic particles interactions**

Primary cosmic rays (CR), i.e. those not produced by interactions with the atmosphere atoms, in the general Solar System environment (e.g. L2 Earth-Sun Lagrangian point) are mostly of galactic origin. They are mostly composed by protons (90 %) and helium nuclei (9 %). The remaining 1% is composed by electrons and heavier nuclei, mostly alphas. The spectrum is peaked at around 200 MeV, and a dominant fraction of such particles is energetic enough to penetrate until the instrument focal plane (e.g. E > 20÷50 MeV). The total flux of the penetrating primaries, integrated over the entire solid angle and the energy, is of the order 2÷5 counts·cm$^{-2}$·s$^{-1}$ depending on solar activity (higher for low solar activity). This figure has been experimentally confirmed in the framework of the Planck mission, whose focal plane was composed by a few tens of individual bolometers.

Since large arrays of LEKID are fabricated on a common bulk Silicon wafer, the number of hits per substrate can be pretty large. The typical arrays that we have employed for NIKA2, and that we will use for B-SIDE, occupy an area of the order of 80 cm$^2$. The rate of hits on the whole substrate is thus expected to be the in order of 160÷400 Hz, which is quite large even considering the fast recovery (response) time of the KID detectors (typically less than a millisecond). The typical energy deposited by a minimum ionizing particle in Silicon is about 500 keV/mm. In order to establish the compatibility of our detectors with the space environment, we have investigated the mechanisms of interaction, transport and generation of spurious signals in large arrays of LEKID sharing a common substrate [8,9].

For the given LEKID detector configuration described in paragraph 1, the incoming particles are not interacting directly with the resonators but rather create phonons in the substrate. These excitations then propagate to several LEKID in the array. At the typical operating temperature, T = 100 mK, and assuming Al ($T_c$ = 1.4 K) and Ti-Al ($T_c$ = 0.9 K), thermal phonons do not have sufficient energy to break Cooper pairs and cannot directly give a measurable signal. The phonons that are detected are thus forcedly non-thermalized. The dominant physical process can be schematized as follows:
- electronic recoils (ionization) from a particle interaction create as a sub-product a cloud of high energy phonons (> 5 THz);
- rapid inelastic scattering results in anharmonic down-conversion until reaching a bottleneck frequency of 1.6 THz (in Silicon) at which point they propagate outward, undergoing elastic scatterings;
- these phonons are above-gap for both Ti-Al and Al superconductors and can thus produce a spurious signal by Cooper-pair breaking;
- the propagation is attenuated mainly by the metallic absorbers (e.g. LEKID film, ground planes, feedline) and the wafer interfaces.

It is worth stressing that the overall process of ionization to Cooper-pair breaking is poorly efficient. We have estimated that in our detectors only a fraction of the order of 0.05% of the energy deposited by the particle is transformed into a detectable signal (quasi-particles in the superconducting film) in the most affected pixel. This behavior is a potential problem for the use of LEKID as high-energy particles detectors. It turns into a significant advantage in the present context of minimization of the CR spurious effect.

**2.3 Impact on standard NIKA detectors**

The standard NIKA 150 GHz detectors are made by Hilbert-shape dual-polarization LEKID [10] with a pixels size of 2.3×2.3 mm$^2$. The arrays used in NIKA1 comprised 132 back-illuminated pixels multiplexed on a single CPW (CoPlanar Waveguide). They are optimized for the band 120÷180 GHz, and are based on Aluminum films with a thickness of 18 nm.

During this experiment, we keep the array under typical space-like illumination conditions, i.e. around 0.5 pW per pixel). The detectors array is irradiated by a $^{241}$Am source placed at a distance of about 600 μm from the central pixel. Collimation is achieved through a hole realized in the λ/4 optical backshort facing the resonators. The $^{241}$Am source produces 5.4 MeV alpha particles. In order to re-scale the absorbed energy to the one corresponding to the peak of the CR spectrum, we set a 10μm Copper shield in front of the source. The resulting alphas exhibit a Gaussian energy distribution centered at 630 keV.

The main results obtained are reported in figure 2. Focusing on the central pixel that is hit by the particle, we show that a single point, in the time-trace acquired at a rate of 200 samples-per-second, is affected. This confirms the fact that LEKID recover their initial state in less than 5 milliseconds. The CR hit also affects significantly a few tens of the surrounding pixels, for an average surface of the order of 1.6×1.6 = 2.6 cm$^2$. Considering these results, and with the aid of a model developed in [6], we conclude that less than 1% of the data would be lost per pixel by a NIKA array placed under space-like conditions. This is to be compared with about 15% for the case of a single Planck bolometer.

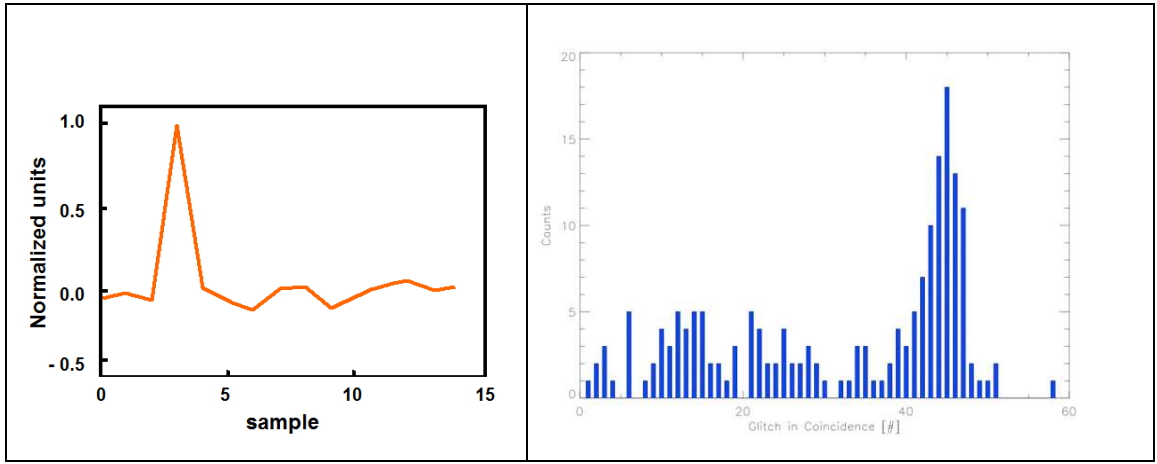

Fig. 2. *Left:* stacked CR hit recorded at a sampling rate of 200 s$^{-1}$. A single sample point is affected. *Right:* distribution of the coincidences around the central detector. An average number of pixels compatible with a 7×7 array around the central one is affected.

## 2.4 Impact on hardened NIKA detectors

As explained in a previous paragraph, the propagation of the non-thermalized phonons in the Silicon substrate can be attenuated using suitable phonon traps. The Aluminum ground plane surrounding the pixels represents the first example of such a trap. In order to downgrade even more efficiently the phonons energy below the Aluminium gap, i.e. harmless phonons for the LEKID, we have added on the back of the wafer a Titanium layer. Titanium, a superconductor with critical temperature around 0.45 K (superconducting gap $\Delta^{Ti}_g < \Delta^{Al}_g$ ), is chosen in order to prevent secondary phonons from reaching the energy of the Aluminum and be able to break Cooper pairs in the LEKID. The proximity of this lossless/superconductor layer does not affect the internal quality factor of the resonators as would be the case for normal metal films like gold. This approach has been first proposed in [11].

This trick prevents the adoption of the classical back-illumination. However, front-illuminated configurations are proven to be equally efficient, and are adopted in operating instruments like for example in NIKA2 [5]. Phonon-traps of this kind might also be implemented, for example, on single-polarization back-illuminated LEKID by replacing the continuum metallic plane by a wire grid.

When tested in the same irradiation/readout conditions described in paragraph 2.3, the NIKA array, back-coated with a Ti layer (hardened), exhibits a rate of coincidences much smaller as shown in figure 3. The size of the array that is affected is now about 0.5×0.5 = 0.25 cm$^2$, a factor ten smaller than in the previous case. The data loss per pixel will be reduced accordingly. This fully confirms the effectiveness of such phonon-trap layer in order to limit the portion of the large array that is affected by a cosmic hit.

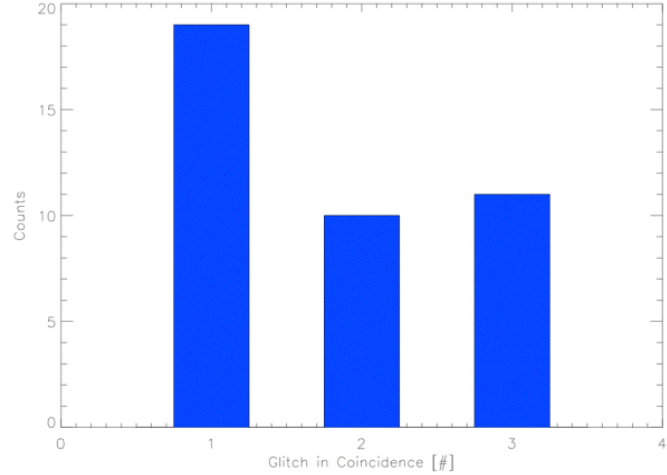

Fig. 3. Distribution of the coincidences around the central detector (same pixel as in fig. 2 for the un-hardened case). An average number of pixels not exceeding a 2×2 array around the central detector is affected.

## 3. THE B-SIDE BALLOON

The next generation of CMB experiments, known as Stage-3 in the US, targets a sensitivity to primordial-gravity-waves-induced polarization B−modes (large-scale "vortexes") expressed in terms of the so-called tensor-to-scalar ratio of r = T/S ≈ 0.01. These instruments, operating from the ground, map precisely the CMB polarization up to frequencies of the order of 220 GHz. At least an equivalent precision should be attained in the removal of foreground emission, more easily detectable at higher frequencies. These higher frequencies, over the large angular scales required, are not accessible from the ground due to atmospheric absorption and noise.

B-SIDE is a proposed balloon-borne experiment designed to map foreground (dust) polarization with high sensitivity and in the band 400÷700 GHz. The main characteristics of the instrument are listed in the table 1.

Table. 1. Summary of the B-SIDE instrument characteristics.

|  | Specifications | Goals |
|---|---|---|
| **Primary mirror diameter (m)** | 0.8 | |
| **Instantaneous field-of-view (deg)** | 2 | 3 |
| **Angular resolution (arc-min)** | 7 | 5 |
| **Number of bands** | 1 | 2 |
| **Flight Duration (days)** | 1 | 3 |
| **Operating frequencies (GHz)** | 450-630 | 400-600 & 500-700 |
| **Number of pixels** | 980 | 1800 |
| **NEP ( W/Hz$^{0.5}$ )** | $5 \cdot 10^{-16}$ | $2 \cdot 10^{-16}$ |
| **Background per pixel** | 50-100 pW | |

### 3.1 Balloon environment

Besides the intrinsic scientific relevance, B-SIDE will be an ideal platform to qualify the LEKID technology, at least for what concerns the impact of cosmic particles, above the atmosphere. The B-SIDE balloon will be flown on a standard

CNES gondola at an altitude of around 40 km surrounded by a residual atmosphere of a few millibars. At altitudes exceeding 20 km, the cosmic particles flux is dominated by the primaries described in a previous paragraph. We will thus conservatively assume encountering the same conditions as in deep space.

The readout electronics do not have to be qualified for long space missions, but we will face new constraints, for example in terms of power dissipation in vacuum environment.

### 3.2 B-SIDE design

The B-SIDE cryostat will employ, for the first time in a real instrument, parts of the space-compatible closed-cycle dilution refrigerator developed at the Institut Néel in view of future satellites (e.g. ATHENA+, CORE+, LiteBird, PIXIE etc.). The expected base temperature will be comprised between 100 and 200 mK, to cool down to a suitable temperature two arrays of LEKID separated by a 45 deg grid polarizer.

At a temperature of about 10 K, a fast-rotating (2÷10 Hz) half-wave plate will ensure the polarization modulation. The half-plate will be realized using the multi-mesh technology [12], and mounted on a superconducting ($MgB_2$, $T_c$ = 40 K) ring to be inserted in a friction-less mechanism based on superconducting magnetic levitation.

The cold optics mixes reflective (M2, M3) and refractive elements. A cold stop at 1 K, and baffles at several temperature stages, strongly suppress the stray light. The lenses, in the baseline option, are realized in Silicon with Anti-Reflective (AR) coating by hot-pressing a suitable intermediate index (n = 1.7÷1.8) dielectric.

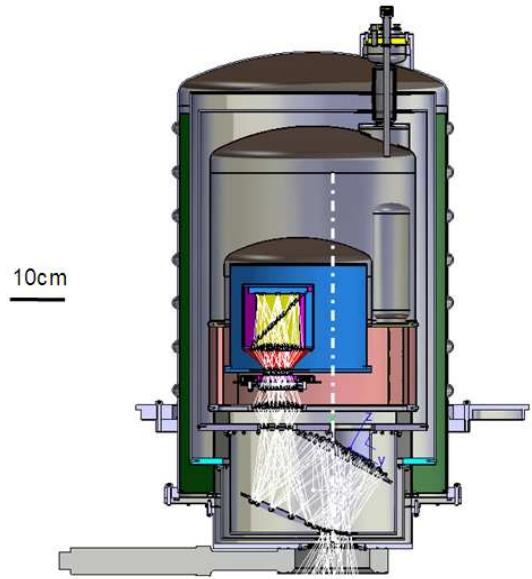

Fig. 4. B-SIDE cryostat cross-section. Preliminary design.

### 3.3 Detectors preliminary performances

The main subsystems of the B-SIDE instrument, e.g. detectors, electronics, half-wave plate mechanism, lenses, cryostat, are under development. In particular, we have designed, fabricated and tested prototype B-SIDE detectors adopting two distinct formats: a) small arrays (60 pixels) to validate the design and test the pixels under irradiation; b) large arrays (490 pixels) already compatible with the B-SIDE specifications. We have demonstrated good sensitivity, preliminarily matching the specification/goals order of magnitude, in the baseline band 450÷630 GHz.

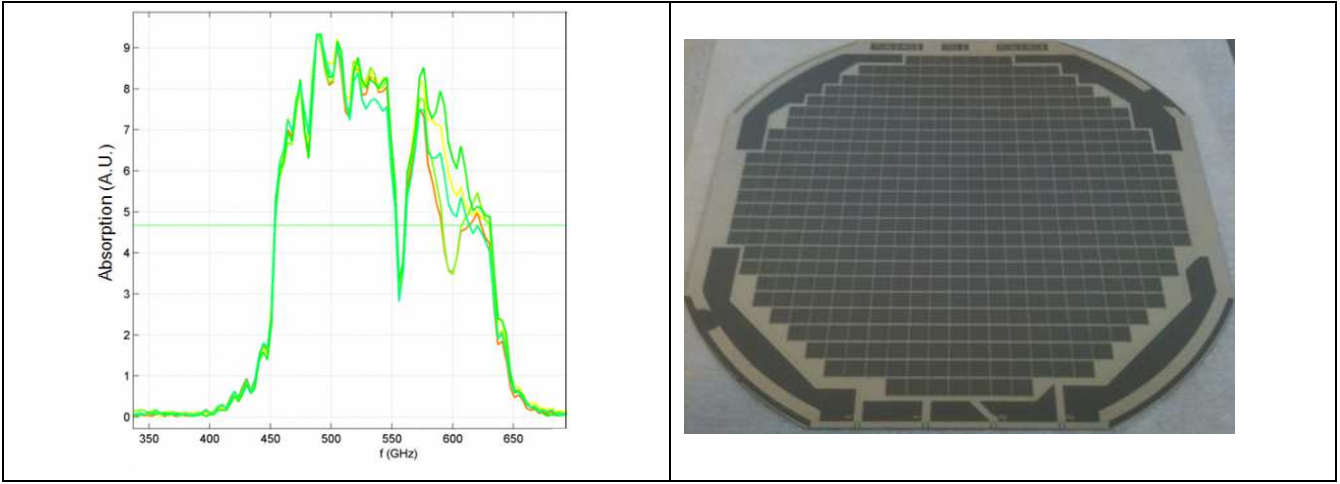

Fig. 5. *Left:* relative spectral characterization of a B-SIDE 60-pixels test array. A few pixels spectra are superposed. The feature at 556 GHz is due to atmospheric absorption. *Right:* a picture of a full 493-pixels B-SIDE array realized on a full 4 inches silicon wafer. On this particular array, 482 out of 493 resonances have been identified, for a ratio of around 98%.

One of the B-SIDE test arrays (60-pixels) has been irradiated with the $^{241}$Am source as described in par. 2.3 and 2.4. These arrays, and the 493-pixels ones, are realized using the classical technology, i.e. without Titanium phonon trap. The number of observed coincidences (an array of 4×4 pixels around the one that is irradiated) confirms the results reported in par. 2.3. In particular, a single sample is affected and the sensitive surface is of the order of 2.6 cm$^2$. We conclude that these detectors are, according to our estimations, compatible with the stratospheric balloon environment. During the R&D associated to the B-SIDE program, we will investigate in parallel front-illuminated configurations allowing implementing the phonons-trap recipe described in paragraph 2.4.

## 4. CONCLUSIONS

We have tested LEKID arrays originally designed for ground-based applications under representative space conditions in terms of background. To study in depth the effect of primary cosmic particles, we have irradiated our arrays using an alpha source, demonstrating good performance, in terms of data loss, even of the classical detectors under such conditions. In addition, we have elaborated a method to further attenuate the effect of the non-thermalized phonons propagating in the substrate.

We are planning to fly large arrays of LEKID on a CNES stratospheric balloon in 2018-19. The design of this instrument, named B-SIDE, has been described briefly. We have produced and optically characterized the first two generation of detectors, showing very promising performance in the band 450÷630 GHz.

## ACKNOWLEDGEMENTS

This work has been partially funded by FP7 program "SPACEKIDS", ANR "NIKA2", CNRS, LabEx "FOCUS" and CNES R&T. We acknowledge the invaluable support from the cryogenics and electronics groups in our laboratories. We acknowledge the support of the micro-fabrication facilities PTA-Grenoble and IRAM-Grenoble. We have had useful discussions with all the members of the NIKA2, SPACEKIDS and B-SIDE collaborations.